# Streptavidin-Biotin Binding is Intrinsically Non-Cooperative, but Experimental Artefacts Can Induce Apparent Positive Cooperativity

Malavika Kayyil Veedu,[1] Eléonore Martenne Duplan,[1] Saurabh Rai,[1] Prithu Roy,[1,#] and Jérôme Wenger[1,*]

[1] *Aix Marseille Univ, CNRS, Centrale Med, Institut Fresnel, AMUTech, 13013 Marseille, France*

** Corresponding author: jerome.wenger@fresnel.fr*

[#] Current address : Swave Photonics, Leuven 3001, Belgium

**Abstract:** The biotin–streptavidin interaction is a key component of nanobiotechnology, yet the nature of its cooperativity remains a debated question. Here, we introduce a label-free spectroscopy approach based on confocal UV autofluorescence microscopy to probe the binding cooperativity of this high-affinity system. By leveraging the intrinsic autofluorescence of tryptophan residues, we avoid artefacts associated with extrinsic fluorescent labels, enabling precise measurements of fully biotin-saturated streptavidin across a wide range of molar ratios. Our experimental results, supported by Monte Carlo simulations, provide conclusive evidence for the absence of cooperativity. However, we show that false signs of apparent positive cooperativity can emerge from local concentration inhomogeneities occurring when rapid reaction kinetics outpace mixing. By establishing the non-cooperative nature of biotin–streptavidin binding and identifying the key conditions to avoid cooperativity artefacts, these insights resolve a longstanding debate and offer critical implications for streptavidin-based biotechnological applications.

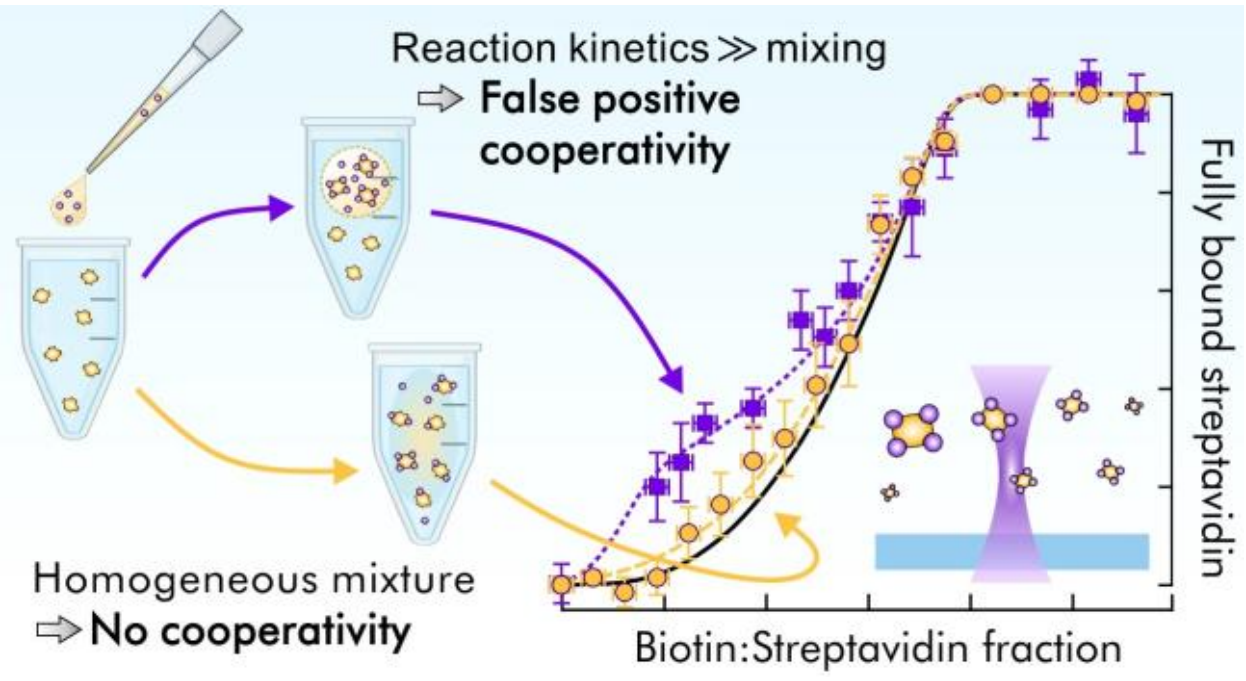

The streptavidin–biotin interaction stands out as a prime example of a biochemical reaction with an extremely high affinity,[1] with its dissociation constant $K_d$ around $4.10^{-14}$ M being frequently reported as one of the strongest noncovalent interactions in biology.[2–4] This exceptional affinity, combined with a remarkably low dissociation rate and a high specificity, has established the biotin−streptavidin complex as a major tool in nanobiotechnology, underpinning major applications in immunoassays, cross-linking and immobilization strategies.[5–10]

Each streptavidin homotetramer holds 4 binding sites for biotin, raising the question of whether streptavidin–biotin interaction exhibits some sort of cooperativity.[11,12] Clarifying the nature of this cooperativity is crucial for nanobiotechnology applications, where precise control over binding kinetics is essential. Despite significant efforts over the last decades, the scientific debate remains open, with studies reporting positive,[13–18] negative,[19–21] or absent cooperativity.[22–24] The lack of scientific consensus in the cooperativity claims underscores the challenge of dissecting the association dynamics governing this ultra-high-affinity interaction.

Early observations of the electrophoretic mobility of the streptavidin-biotin complex in presence of urea suggested positive cooperativity.[13] However, chromatographic separation of the streptavidin-biotin complexes after 2 days equilibrium revealed similar affinities between binding sites and concluded to the absence of cooperativity.[22] Although this work by Jones and Kurzban is often cited to indicate that the affinity per binding site remains essentially unchanged as the fraction of bound sites increases,[22] it is important to note that these findings were obtained under equilibrium conditions reached after 2 days. Under such prolonged timescales, slow dissociation allows for re-equilibration between binding sites of different neighbouring proteins. However, most modern applications of biotin-streptavidin binding do not reach such extended equilibrium times. The tight binding of biotin and the rapid association times complicate the analysis and can lead to non-equilibrium conditions that appear stable over the duration of typical experiments. Therefore, it is critical to probe the biotin-streptavidin association in conditions most relevant for current biotechnology assays.

More recently, clear evidence has emerged of cooperativity among the 5 hydrogen bonds involved in the binding of a biotin molecule inside a single streptavidin receptor.[4,25–28] Biotin binding was found to induce conformational changes spreading across the different streptavidin subunits and stabilize the complex.[14,29–31] Furthermore, biotin binding has been shown to drastically alter the dynamics of surface residues across the whole protein.[18,32] These observations have been conjected to promote the association of subsequent biotins,[14,18] a phenomenon sometimes referred to as cooperative allosterism.[16] However, it remains unclear whether these effects affect the affinity for the other binding sites, as quantitative proofs of change of the binding affinities are still lacking.

With its single molecule sensitivity and broad dynamic range spanning from pico- to milliseconds, fluorescence microscopy appears ideally suited to probe the biotin-streptavidin reaction kinetics.[33–37] However, the use of fluorescent labels introduces two critical issues. First, Förster resonance energy transfer (homo-FRET) occurs between neighbouring fluorophores owing to the nanometer separation between the streptavidin subunits. This leads to a significant quenching of the signal which complicates the analysis of the binding events.[21,35,38] Second, the presence of fluorescent labels may alter the protein conformation and/or its reaction dynamics.[39–43] In the case of streptavidin, dye coupling has been shown to significantly impair the binding of biotin,[44] leading to the observations of accelerated ligand dissociation, decreased ligand association, and reduced specificity.

Here, we introduce a label-free protocol specifically designed to probe cooperativity in the biotin–streptavidin interaction. To harness the high sensitivity of fluorescence microscopy while avoiding artefacts associated with extrinsic fluorescent labels, all experiments were conducted under UV excitation at 295 nm to directly probe the intrinsic autofluorescence of tryptophan residues in label-free streptavidin.[45–48] Using a confocal UV microscope system with a calibrated detection volume ensures the high reproducibility required to analyze data across a wide range of biotin-to-streptavidin ratios.[37] Our experimental data, supported by Monte Carlo simulations, provide conclusive evidence for the absence of cooperativity in the biotin–streptavidin interaction. However, we demonstrate that false signs of apparent positive cooperativity can arise from experimental artefacts, particularly when rapid reaction kinetics outpace mixing and lead to local concentration inhomogeneities. By establishing the non-cooperative nature of biotin–streptavidin binding and identifying the key conditions to avoid cooperativity artefacts, our findings resolve a longstanding debate and pave the way for more accurate and reliable biotechnological applications based on streptavidin.

Figure 1a schematically outlines our methodology designed to investigate the cooperativity in the biotin–streptavidin interaction. The core idea is to record the amount of fully bound streptavidin proteins holding 4 biotins as the biotin to streptavidin molar fraction is increased. The streptavidin concentration is held constant at 200 nM, while varying concentrations of biotin are introduced into the solution. Following incubation, the non-fully bound proteins with 1, 2, 3 or 4 unoccupied receptor sites are filtered from the solution using biotin-coated silica beads of 1 µm diameter. The relatively large size and density of these beads enable efficient separation by centrifugation at 7,000 rpm for less than 2 min. Finally the concentration of fully-bound streptavidin remaining in the solution is measured using a confocal UV microscope set to record the intrinsic autofluorescence of the streptavidin tryptophan residues under 295 nm excitation with high sensitivity and low background.[48,49] No extrinsic fluorescence labelling is involved in our experiments. All our measurements are performed in solutions to avoid the steric hindrance and electrostatic repulsion introduced by surfaces.[19] Continuous stirring of the solution during the UV measurements combined with the use of a 100 mM potassium phosphate buffer (pH 6.6) prevents from observing photobleaching or UV-induced photodamage, as shown in Fig. S1. This active stirring was only applied during the UV autofluorescence measurements to prevent photodamage. The initial mixing of biotin and streptavidin was performed by pipetting the biotin solution into the streptavidin sample followed by vortex stirring.

Our main results are typically presented as a binding curve as shown in Fig. 1b, which plots the fraction of fully bound (biotin-saturated) streptavidin as a function of the biotin-to-streptavidin molar ratio. As each streptavidin bears 4 biotin binding sites, we expect full saturation of all streptavidin proteins at biotin-to-streptavidin molar ratios exceeding 4, leading to a plateau in the binding curve. The evolution of the fully bound streptavidin fraction at intermediate biotin-to-streptavidin molar ratios provides insight into the cooperativity of the binding process.

While classical cooperativity models are typically used to describe the average ligand occupancy per molecule,[50] they are not directly applicable to our experimental setup. Due to the similar optical response of streptavidin molecules with 0, 1, 2, 3, or 4 bound biotins (Fig. S4), UV autofluorescence cannot directly resolve fractional occupancy. Instead, our protocol is specifically designed to monitor only fully saturated streptavidin molecules, which forms a measurable quantity in our UV microscope. For this reason, we employed Monte Carlo simulations, which allow us to model the experimental binding curve under these specific conditions. The complete Matlab code used for the simulations is detailed in the Supporting Information Section S3.

Monte Carlo simulations based on repeated trials allow us to predict the evolution of the binding curve for different cooperativity scenarios. We introduce a cooperativity parameter $p$ for our calculations, where $p = 1$ represents mutually independent binding events (non-cooperativity), $p > 1$ stands for positive cooperativity, while $p < 1$ corresponds to negative cooperativity. The probability of a biotin molecule binding to a specific streptavidin protein is weighted according to the relative population $N_i/N_{tot}$ of streptavidin molecules bearing a number $i$ of bound biotins, multiplied by a supplementary term depending on the cooperativity parameter $p$. We explored multiple models for this cooperativity term, all yielding qualitatively similar results such that the model choice does not affect our primary conclusions. Explicit expressions are provided in the Supporting Information Section S3 and Fig. S2.

In the case of strong positive cooperativity, biotin molecules preferentially saturate a subset of streptavidin proteins, leading to a steep increase in the binding curves shown in Fig. 1b and S2. Conversely, under strong negative cooperativity, the full saturation of all 4 binding sites becomes highly unfavourable, resulting in a very slow increase in the binding curve. In the absence of cooperativity, the binding curve follows a smooth gradual increase (black trace in Fig. 1b), which is derived from Monte Carlo simulations assuming mutually independent binding events. The main goal of our experiments is to record the fraction of fully biotin-saturated streptavidin and compare the data evolution to the simulations in Fig. 1b to determine the nature of the cooperativity (positive, negative, absence).

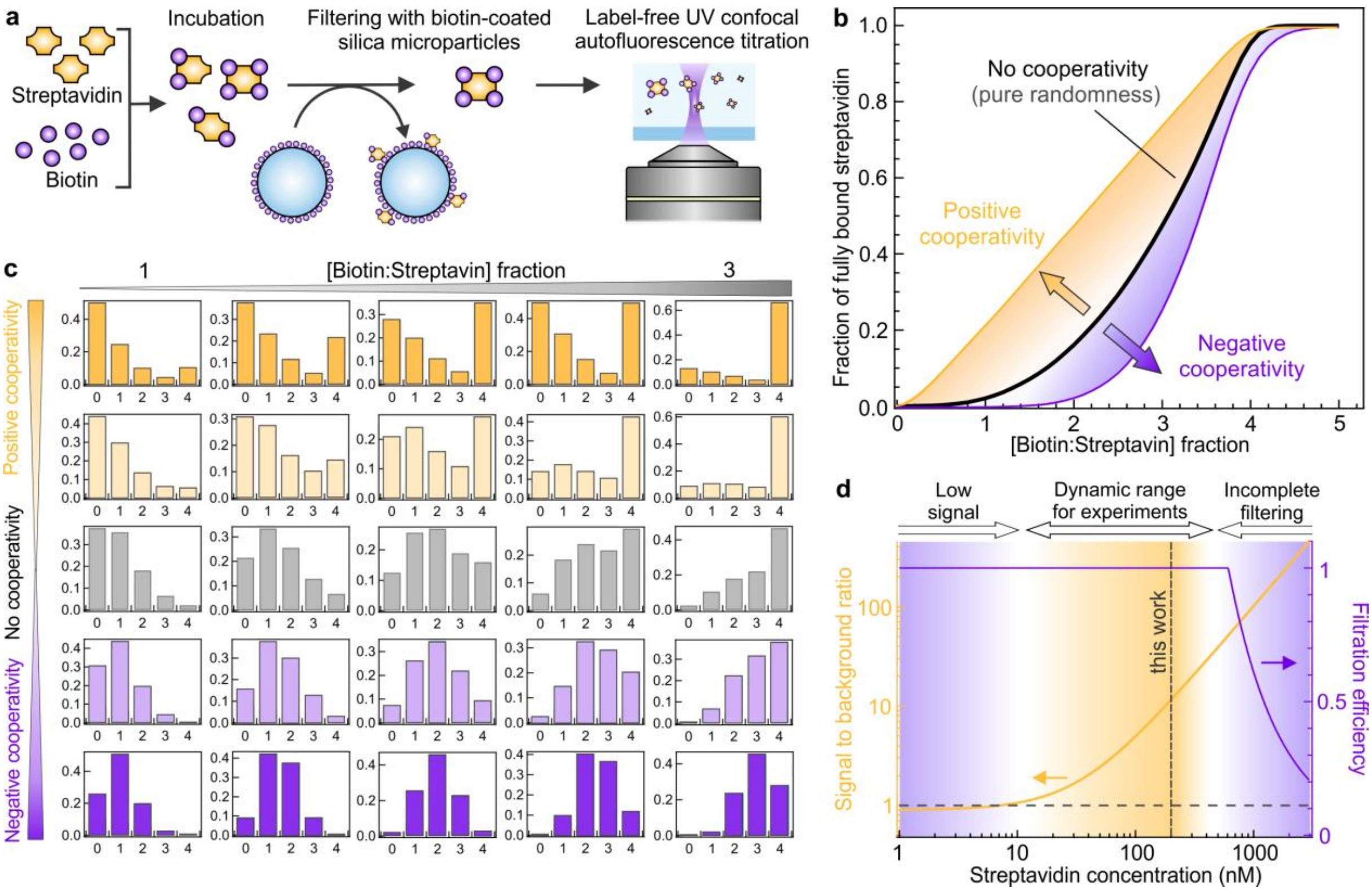

**Figure 1.** Concept to probe cooperative binding for the biotin-streptavidin system. (a) Scheme of the experimental protocol to measure the fraction of fully bound streptavidin proteins. (b) Monte Carlo simulations of the fraction of fully bound streptavidin molecules bearing 4 biotins as a function of the biotin to streptavidin relative concentration. The thick black trace corresponds to the absence of cooperativity (mutually independent binding events). The yellow and purple regions correspond respectively to different values of positive and negative cooperativities (see Supplementary Information Fig. S2 for simulations using different models for cooperative binding, the conclusions remain similar among the different models used). (c) Each subgraph represents the relative fractions of streptavidin molecules with 0, 1, 2, 3 or 4 bound biotins. The columns correspond to a specific ratio of biotin to streptavidin concentrations with increments of 0.5 from 1 on the left to 3 on the right. The lines correspond to a specific cooperativity condition, from highly negative at the bottom to strongly positive at the top. (d) Determination of the suitable experimental range. The orange curve (left axis) represents the signal to background ratio (SBR) as a function of the streptavidin concentration for the UV autofluorescence measurement. Here we have taken the background as the blank reference value in absence of streptavidin plus 9 times the standard deviation to account for the limit of quantitation.[37,51] The SBR value must exceed 1 for the experiment to remain quantitative. The purple curve (right axis) shows the predictions of the filtration efficiency for our biotin-coated silica beads, which decreases when concentrations higher than 600 nM of streptavidin are used.

Figure 1c helps to visualize and compare the simulated distributions of the relative populations of streptavidin bound to 0, 1, 2, 3, or 4 biotins. From left to right, the biotin-to-streptavidin fraction increases, while from top to bottom the cooperativity goes from highly positive to highly negative. As the biotin-to-streptavidin ratio increases, we observe a gradual progression from lower to higher occupancy states. Under positive cooperativity, biotin molecules preferentially saturate streptavidin proteins, leading to a higher fraction of fully bound proteins even at moderate biotin-to-streptavidin ratios. Conversely, negative cooperativity results in a more uniform distribution of partially occupied proteins, delaying the emergence of fully saturated streptavidin.

Beyond the label-free nature of our measurement, the experimental challenges arise from a trade-off between sensitivity and filtration efficiency, as shown in Fig. 1d. The titration of the fully bound streptavidin concentration requires enough signal to exceed the background level, which is typically achieved on our setup for streptavidin concentration above 10 nM. While increasing the streptavidin concentration improves the signal strength, it simultaneously demands greater filtration efficiency to effectively remove all the non-fully-saturated streptavidin. We have tried commercially-available biotin-coated microparticles (Sigma Aldrich ref. 79195), but their filtration efficiency was inadequate for our purpose. Instead, we have developed custom biotin-coated silica beads, with details provided in the Methods Section. Using our tailored system, the filtration efficiency is predicted to drop for streptavidin concentrations exceeding 600 nM (blue trace in Fig. 1d, right axis). To ensure robustness, we conducted our experiments at a streptavidin concentration of 200 nM, providing a suitable safety margin. The Supporting Information Figure S3 demonstrates experimentally that our custom biotin-coated silica beads are sufficient to filter all the streptavidin proteins, resulting in a non-detectable bound fraction of streptavidin in the absence of biotin.

Our protocol requires the experimental calibration of the autofluorescence signal as a function of the streptavidin concentration (Fig. 2a). This curve allows us to subsequently determine the streptavidin concentration from the measured autofluorescence intensity after incubation with biotin and filtration. The presence of biotin tends to reduce both the autofluorescence brightness and lifetime by about 36% (Fig. S4), in consistency with an earlier report.[52] We take this effect into account by performing both the calibration and experimental detection on fully biotin-saturated streptavidin.

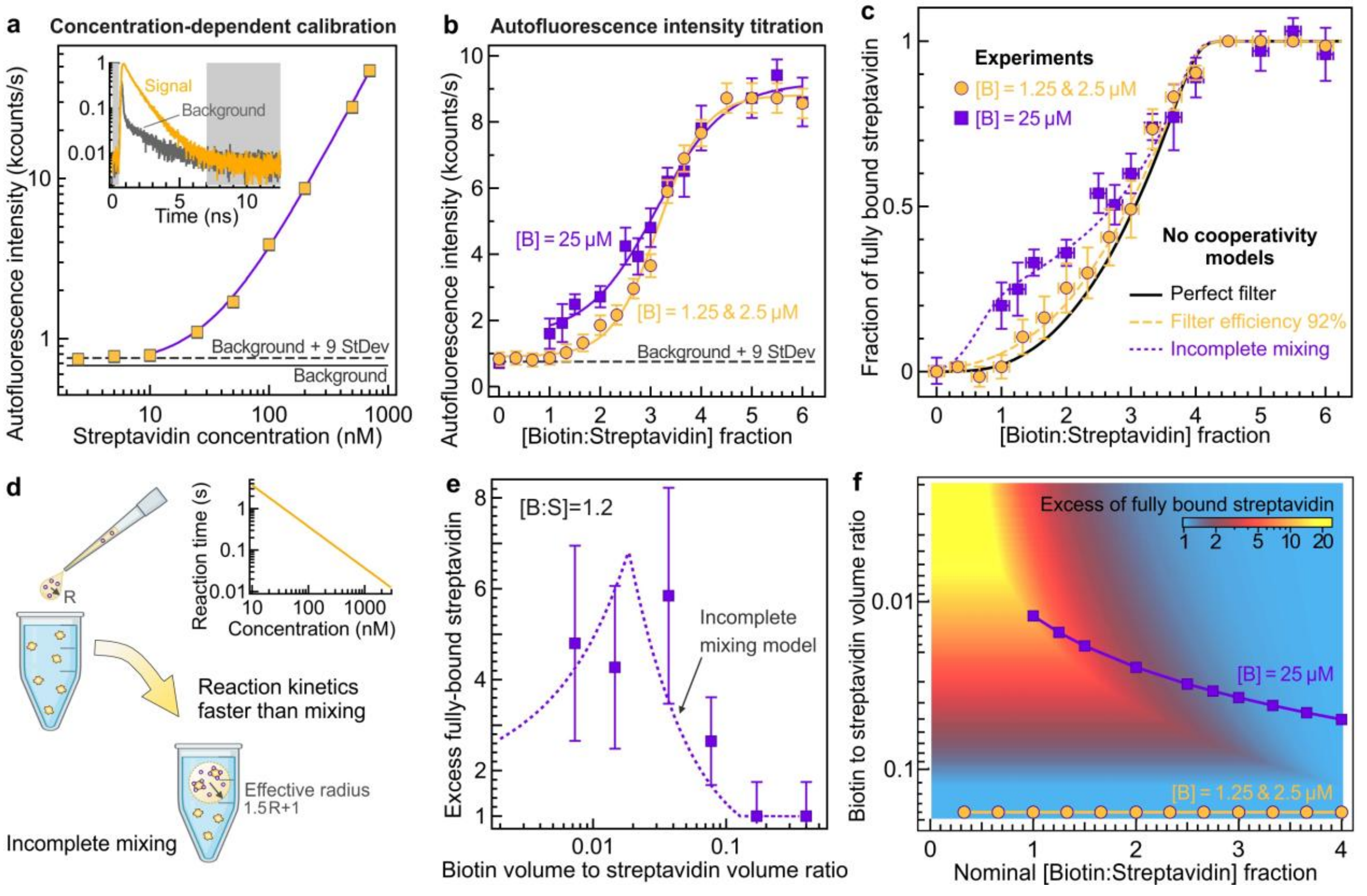

**Figure 2.** Experiments show no cooperativity for the biotin-streptavidin association, but incomplete mixing may falsely suggest positive cooperativity. (a) Calibration of the UV autofluorescence intensity for different known concentrations of label-free streptavidin fully saturated by biotin. The horizontal lines represent the background intensity and the background intensity plus 9 times the standard deviation recorded for the blank reference in absence of streptavidin.[51,53] The insert shows the 0.5 to 7 ns time gate applied to the autofluorescence lifetime histogram in order to improve the signal-to-background ratio.[48,54] (b) Experimental autofluorescence intensity measured after applying the protocol in Fig. 1a for different biotin concentrations. Purple markers refer to experiment performed using a stock biotin concentration of 25 µM. Orange markers correspond to biotin stock concentration of 1.25 µM for B:S ratios below 2 and 2.5 µM for B:S ratios higher than 2. The curves are sigmoidal fits to the data. (c) Fraction of fully bound streptavidin as a function of the biotin to streptavidin relative concentration. The experimental data correspond to the measurements in (b) combined with the calibration in (a). The black curve corresponds to the Monte Carlo predictions in the absence of cooperativity for a perfect filtration of non-fully saturated streptavidin. The dashed orange curve represents simulations in absence of cooperativity with 92% filtration efficiency. The dotted purple curve shows the simulations results in absence of cooperativity taking into accounts the effects of incomplete mixing at high biotin concentration (see text for details). (d) Model to estimate the influence of incomplete mixing: a biotin solution droplet of volume $\frac{4}{3}\pi R^3$is added to the streptavidin

sample. As shown in the insert show, the reaction kinetics can be significantly faster than the typical mixing time. This leads to an effective interaction volume of radius $1.5R + 1$, where all distances are expressed in mm. (e) Measurement of the excess of fully-bound streptavidin as a function of the ratio between the biotin and streptavidin volumes. The molar fraction between biotin and streptavidin is kept constant at 1.2. The dotted line shows the predictions from the model in (d) with no cooperativity. (f) Bidimensional map of the excess of fully-bound streptavidin as a function of the ratio between the biotin and streptavidin volumes and the biotin to streptavidin relative concentration (assuming perfect mixing). The calculations use the incomplete mixing model in (d) validated experimentally in (c) and (e). No cooperativity is involved in these simulations. The markers indicate the conditions of the experiments in (b,c). In (b,c,e), the error bars correspond to 2× the standard deviation.

Following the protocol illustrated in Fig. 1a, we incubated a 200 nM streptavidin solution with varying biotin concentrations, then filtered out non-fully bound proteins before measuring the autofluorescence intensity of the fully saturated streptavidin. Two experimental datasets are presented in Fig. 2b, distinguished by the biotin stock concentrations used for dilution: a high-concentration case (25 µM, purple markers) and a lower-concentration set (1.25 µM for biotin-to-streptavidin ratios below 2 and 2.5 µM for ratios above 2, orange markers). For a 200 µL incubation volume, the 25 µM biotin stock corresponds to typical biotin volumes of a few microliters, whereas the lower concentrations ensure biotin volumes always exceed 15 µL. These concentration and volume choices are critical, as discussed later. Notably, the raw intensity data in Fig. 2b already reveal clear differences for biotin-to-streptavidin molar fractions below 3, depending on the biotin stock concentration used during mixing, although all other experimental conditions remain identical.

The combination of the autofluorescence intensity data from Fig. 2b with the calibration curve in Fig. 2a determines the experimental fraction of fully bound streptavidin as a function of the biotin-to-streptavidin molar ratio (Fig. 2c), which represents our primary result. The two datasets exhibit distinct behaviors: the low-concentration case (1.25 and 2.5 µM biotin) aligns closely with the non-cooperative model (black trace), whereas the 25 µM biotin data clearly falls within the positive cooperativity region. Additionally, the saturation of the binding curve for biotin-to-streptavidin ratios exceeding 4 supports the consistency of our findings, in agreement with the known tetrameric structure of streptavidin.[1] As our experiment assesses the fraction of fully saturated streptavidin but cannot resolve the fractional occupancy of the total receptor sites as a function of the ligand concentration, we cannot extrapolate our data to draw the Hill plot frequently used in the context of cooperativity.[12]

Within the experimental uncertainties, the orange data in Fig. 2c for low biotin concentrations align closely with the non-cooperative model assuming mutually independent binding events (thick black trace). This experimental dataset can be fitted with an improved accuracy by accounting for some incomplete filtration of non-fully saturated streptavidin depending on its ligation state, while still assuming non-cooperative binding. Empirically, we found that filtration efficiencies of 100%, 96%, 92%, and 92% for streptavidin proteins with 0, 1, 2, or 3 bound biotins, respectively lead to an excellent agreement with the experimental results (dashed orange line in Fig. 2c). Importantly, this demonstrates that a simple model not invoking any cooperativity can fully explain these experimental observations. Alternatively, an equivalent non-cooperative model assumes that a small fraction of streptavidin proteins lack fully active binding sites. In this case, fitting the low biotin concentration dataset would require assuming that 4%, 8%, and 8% of streptavidin molecules miss 3, 2, or 1 binding sites, respectively, leading to a similar outcome as the dashed orange line in Fig. 2c. While we cannot currently distinguish between these two interpretations, the excellent agreement of the data for low biotin concentrations with these realistic non-cooperative models strongly supports the absence of cooperativity in the streptavidin-biotin interaction at moderate biotin concentrations.

If the streptavidin-biotin interaction is non-cooperative, why do the experimental data for 25 µM biotin concentration (purple dots in Fig. 2c) clearly fall into the positive cooperativity region? As discussed below, at these high biotin concentrations, the small biotin volumes of a few microliters combined with rapid reaction kinetics lead to local concentration inhomogeneities, which in turn create apparent cooperativity artefacts.

Calculations of the reaction times show that at micromolar concentrations, the association between biotin and streptavidin occurs within a few tens of milliseconds, much faster than typical mixing times (insert of Fig. 2d). This estimation uses the experimentally determined biotin-streptavidin association rate constant $k_{ON}$ of 2.7 $10^7$ $M^{-1}s^{-1}$.[37] At high biotin concentrations, the rapid association can outpace mixing, resulting in concentration gradients in the mixture. To address these inhomogeneities, we introduce the model illustrated in Fig. 2d, which assumes that the interaction occurs within a spherical volume of radius $1.5R + 1$, where $R$ is the radius of the biotin droplet and all distances are expressed in mm. The slope of 1.5 was derived from experiments visualizing the early dilution stages of black ink droplets of µL volumes in water (Supporting Information Fig. S5). The motivation for these experiments was to determine how the effective interaction volume scales with the droplet radius under realistic experimental conditions. The +1 mm offset was determined empirically by fitting the 25 µM biotin data in Fig. 2c (see further discussion in the Supporting Information section S7).

The effective interaction volume smaller than the total solution volume locally increases the biotin-to-streptavidin molar ratio, promoting the saturation of a subset of streptavidin proteins. The results of this “incomplete mixing” model, which involves no cooperativity, are shown as the dotted purple curve in Fig. 2c, closely matching the 25 µM biotin dataset. Importantly, this agreement demonstrates that the 25 µM data points, which appear to fall well into the positive cooperativity region, can be fully explained by incomplete mixing without invoking cooperativity. Improper choice of biotin concentration and droplet volume can create concentration inhomogeneities when reaction kinetics outpace the mixing kinetics. This can potentially lead to misinterpretations of positive cooperativity where none exists.

While our results do not address potential variations in the binding affinities of individual sites at different levels of biotin saturation, they demonstrate that the global interaction dynamics of streptavidin remain unchanged as successive biotin-binding sites become occupied. Upon the binding of biotin, the streptavidin structure can become more compact and get stabilized,[14,16] characterized by enhanced thermal stability,[29] and reduced hydrophobicity.[31] However, our results show that these independent observations do not significantly promote the association of subsequent biotins, resulting in an absence of cooperativity for the biotin-streptavidin binding. While comparison with cooperative models (Fig. S2) does not allow us to completely rule out minor positive cooperativity, the strong dependence of our results on biotin concentration, combined with the ability of simple realistic non-cooperative models to fully explain the observations, indicates that cooperativity does not play a significant role in the streptavidin–biotin interaction.

To further validate our model, we have conducted experiments at a constant biotin-to-streptavidin molar ratio of 1.2, where the effects of incomplete mixing and rapid reaction kinetics are most pronounced. We have varied the biotin stock concentrations, resulting in different biotin-to-streptavidin volume ratios. The excess of fully bound streptavidin was calculated relative to a reference recorded at a biotin-to-streptavidin volume ratio of 0.4 (Fig. 2e). In this reference case, the large volumes ensure rapid mixing and prevent concentration inhomogeneities. As the biotin-to-streptavidin volume ratio decreases, the excess of fully bound streptavidin increases due to the high local biotin concentrations and small volumes involved, consistent with the "incomplete mixing" model in Fig. 2d. Our simple non-cooperative model effectively reproduces the experimental observations (dotted curve in Fig. 2e). Although dividing by small numbers increases the error bars, the core trend remains evident. The peak observed at a biotin-to-streptavidin volume ratio of 0.02 corresponds to an effective biotin-to-streptavidin molar ratio of 4 within the interaction volume, maximizing the number of fully saturated proteins.

Having validated our model with the data in Fig. 2c and 2e, we extend the analysis to a range of biotin-to-streptavidin molar fractions and volume ratios (Fig. 2f). This two-dimensional map serves as a global guideline to predict conditions that may lead to local excesses of saturated proteins, resulting in artefacts resembling positive cooperativity. The blue regions in Fig. 2f represent safe zones, free from concentration gradients under typical experimental conditions. In particular, using biotin-to-streptavidin volume ratios above 0.15 prevents from observing concentration inhomogeneities. In stark contrast, the orange-yellow areas indicate strong concentration inhomogeneities, producing an excess of fully bound streptavidin by more than fourfold. This effect is most pronounced for biotin-to-streptavidin molar fractions below 1 and volume ratios below 0.05. When we map the experimental conditions from Figs. 2b and 2c onto this 2D plot, the 1.25 and 2.5 µM biotin dataset falls within the artefact-free region, while the 25 µM biotin dataset enters the artefact-prone zone at low biotin-to-streptavidin molar fractions.

In summary, we have addressed a longstanding debate surrounding cooperativity in the biotin–streptavidin interaction using a label-free spectroscopy approach based on confocal UV autofluorescence microscopy. By carefully optimizing the signal-to-background ratio, filtration efficiency, and dose-response calibration, we ensured the robustness of our measurements, allowing us to probe the binding curve of fully biotin-saturated streptavidin as a function of the biotin-to-streptavidin molar ratio. Our experimental results, supported by Monte Carlo simulations, provide conclusive evidence for the absence of cooperativity in the biotin–streptavidin interaction. While structural and dynamical changes in streptavidin conformations upon biotin binding have been previously documented,[14,29–31] our data demonstrate that these changes do not translate into cooperative binding behavior under application-relevant conditions. Instead, we show that experimental artefacts, such as incomplete mixing, can induce apparent signs of positive cooperativity, particularly at biotin-to-streptavidin molar fractions between 1 and 3. This effect arises when rapid reaction kinetics outpace mixing, leading to local concentration inhomogeneities that falsely suggest cooperative behavior. These insights resolve a decades-long controversy and underscore the importance of experimental design in accurately interpreting binding dynamics. By clarifying the non-cooperative nature of biotin–streptavidin binding, our findings provide clear guidelines for the rational design and optimization of streptavidin-based biotechnological applications, where precise control over binding kinetics is essential.

**Supporting Information**

Materials and Methods; Homogenizing the solution sample with a rotating stirrer; Matlab code used for numerical simulations; Equal, additive and multiplicative models used for cooperativity; Control of filtration efficiency; Influence of biotin on streptavidin UV autofluorescence; Dimension of the interaction volume; Calibration of the UV confocal detection volume; Convergence check for the simulations; Analytical expression for the mutually-independent binding model.

**Funding**

This project has received funding from the European Research Council (ERC) under the European Union's Horizon 2020 research and innovation programme (grant agreement 723241), from the European Research Executive Agency (REA) under the Marie Skłodowska-Curie Actions doctoral network program (grant agreement 101072818), from the European Union under Horizon Europe Eureka Eurostars program (grant agreement E!8928), and from the Agence Nationale de la Recherche (project ANR-24-CE42-0846, ANR-25-CE42-6199-01).

**Conflict of Interest**

The authors declare no conflict of interest.

**Data Availability Statement**

The data that support the findings of this study data are available from the corresponding author upon request.

## Supplementary Information for

# Streptavidin-Biotin Binding is Intrinsically Non-Cooperative, but Experimental Artefacts Can Induce Apparent Positive Cooperativity

Malavika Kayyil Veedu,[1] Eléonore Martenne Duplan,[1] Saurabh Rai,[1] Prithu Roy,[1] and Jérôme Wenger[1,*]

*[1] Aix Marseille Univ, CNRS, Centrale Med, Institut Fresnel, AMUTech, 13013 Marseille, France*

** Corresponding author: jerome.wenger@fresnel.fr*

**Contents:**

## S1. Materials and Methods

**Samples.** Biotin was obtained from Sigma–Aldrich (ref. B4501) and used without further purification. Streptavidin, derived from Streptomyces avidinii, was also purchased from Sigma–Aldrich (ref. S4762) and stored at −20 °C. A stock solution of 3 µM streptavidin was made in 100 mM potassium phosphate buffer (pH 6.6) with 10 mM NaCl, and a stock solution of 25 µM biotin was made in water. Silica beads of 1 µm diameter were obtained from Sigma-Aldrich (ref. 904457). Silane-PEG-Biotin with a molar mass of 2 kDa was obtained from Nanocs (ref. PG2-BNSL-2k).

To prepare the biotin-coated silica beads used for the filtration step, 50 mg/mL of silica beads prepared in ethanol (98 %) with 0.5 % Tween20 (Sigma Aldrich P1379) was mixed with 5 mg/mL of silane-PEG biotin prepared in ethanol (98 %) with 1 % acetic acid. This sample was incubated overnight with gentle mixing. After incubation, the sample was washed and filtered by centrifugation 1x with ethanol and then 5x with water to remove the unbound PEG-biotin molecules.

**Incubation and filtration protocols.** For each measurement, the target concentration of streptavidin was kept constant at 200 nM. Biotin was added to the protein solution as per the required ratio for measurement with immediate mixing to homogenize the solution. The 200 µL streptavidin-biotin mixture was incubated for 15 min with mixing to ensure maximum biotin binds to streptavidin. 15 min was considered to be largely enough, considering the fast association rate of streptavidin and biotin.[1,2] 100 µL of the biotin-coated silica bead solution was added to the streptavidin-biotin mixture and incubated for 30 min with mixing to prevent the settling of silica beads. This ensures that all the streptavidin that is not fully bound with the added free biotin binds to the biotin on the surface of the silica bead. After incubation, we centrifuged the mixture to separate the silica-biotin beads from the fully bound streptavidin solution and kept only the supernatant. This way we separate the fully bound streptavidin from the streptavidin with one or more vacant biotin binding sites. The final measured solution only contains fully bound streptavidin in 100 mM potassium phosphate buffer and 10 mM NaCl. During our measurement, we did not detect any sign of the biotin-coated silica beads (expected to result in an intense autofluorescence peak) demonstrating the efficiency of the filtration protocol. Figure S3 of the Supporting Information provides supplementary proof for the filtration efficiency in our experimental conditions. The presence of biotin alone did not lead to any detectable UV autofluorescence signal.

**Filtration capacity calculation.** Filtration is a crucial step in our measurement. It is important to ensure that all the streptavidin with vacant biotin binding sites can be removed by the biotin-coated silica bead solution so that only the fully bound biotin remains. At 50 mg/mL, the 1 µm silica bead concentration is $3.61 \times 10^{10}$ beads per mL, which implies $3.61 \times 10^{9}$ beads in 100 µL. Assuming a $2 \times 2$ $nm^2$ density per silane-PEG-biotin molecule, a 1 µm bead requires $7.85 \times 10^{5}$ silane-PEG-biotin molecules. At a coating concentration of 5 mg/mL and a bead concentration of $3.61 \times 10^{10}$ beads per mL, the number of silane-PEG-biotin molecules per bead exceeds $4.1 \times 10^{7}$, thereby largely covering all the bead surface. Considering the streptavidin size, the surface density of streptavidin after coating with the silica-biotin bead is assumed to be around 1 per $10 \times 10$ $nm^2$. Therefore, around 31000 streptavidins can be bound per 1 µm bead. Hence, $3.61 \times 10^{9}$ beads can capture around $1.1 \times 10^{14}$ streptavidin molecules. This corresponds to a maximum streptavidin concentration of 600 nM for a 300 µL solution (assuming all streptavidin can bind to the surface of the beads). To leave a 3x margin for security, we decided to work at 200 nM and 300 µL.

**UV confocal microscope**. The autofluorescence measurements were performed on a custom-built confocal microscope with a LOMO 58x, 0.8 NA, water immersion objective. A 295 nm picosecond laser (Picoquant VisUV-295-590, 70 ps pulse duration, 80 MHz repetition rate) is used for excitation with 6 µW power measured at the entrance port of the microscope. The laser beam is spatially filtered with a 50 µm pinhole and spectrally filtered by a short-pass filter (Semrock FF01-311/SP-25). The excitation laser is then reflected to the sample by a dichroic mirror (Semrock FF310-Di01-25-D). A rotating stirring unit described in the Supplementary Fig. S1 was used to homogenize the solution and mitigate the UV photodamage of the proteins. The fluorescence emission is collected back by the same microscope objective and separated from the excitation by the dichroic mirror and two emission filters (Semrock FF01-300/LP-25 and Semrock FF01-375/110-25). A 200 mm focal length air-spaced achromatic doublet is used as a microscope tube lens to focus the fluorescence light on a 50 µm confocal pinhole. The emission in the spectral range 310 to 410 nm is detected by a single photon counting photomultiplier tube (Picoquant PMA 175) connected to a photon counting module (Picoquant Picoharp 300 with time-tagged time-resolved mode).

The UV autofluorescence measurements were performed immediately after the silica bead filtration and repeated at least 3 times with typically 60 s integration time per measurement. To minimize surface-related artefacts, we initially passivated silica coverslips with 2 kDa PEG-silane (Nanocs PG1-SL-2k). However, no improvement was observed, and we proceeded with untreated pure silica coverslips. At the experimental pH of 6.6, both streptavidin (isoelectric point 5–6)[3] and silica (point of zero charge <4)[4] carry negative charges, reducing electrostatic interactions. Additionally, all measurements were performed in solution, approximately 20 µm away from the surface, further mitigating potential surface effects.

All the data analysis was performed using Symphotime64 software (Picoquant). A time gating filter was applied to the autofluorescence data in order to improve the signal to background ratio by rejecting the photons whose arrival time fall outside the autofluorescence signal window of 6.5 ns duration.[5,6] The confocal detection volume is calibrated to 0.9 fL using a 90 nm nanoaperture milled into an opaque aluminum film (Fig. S6).[7]

## S2. Homogenizing the solution sample with a rotating stirrer

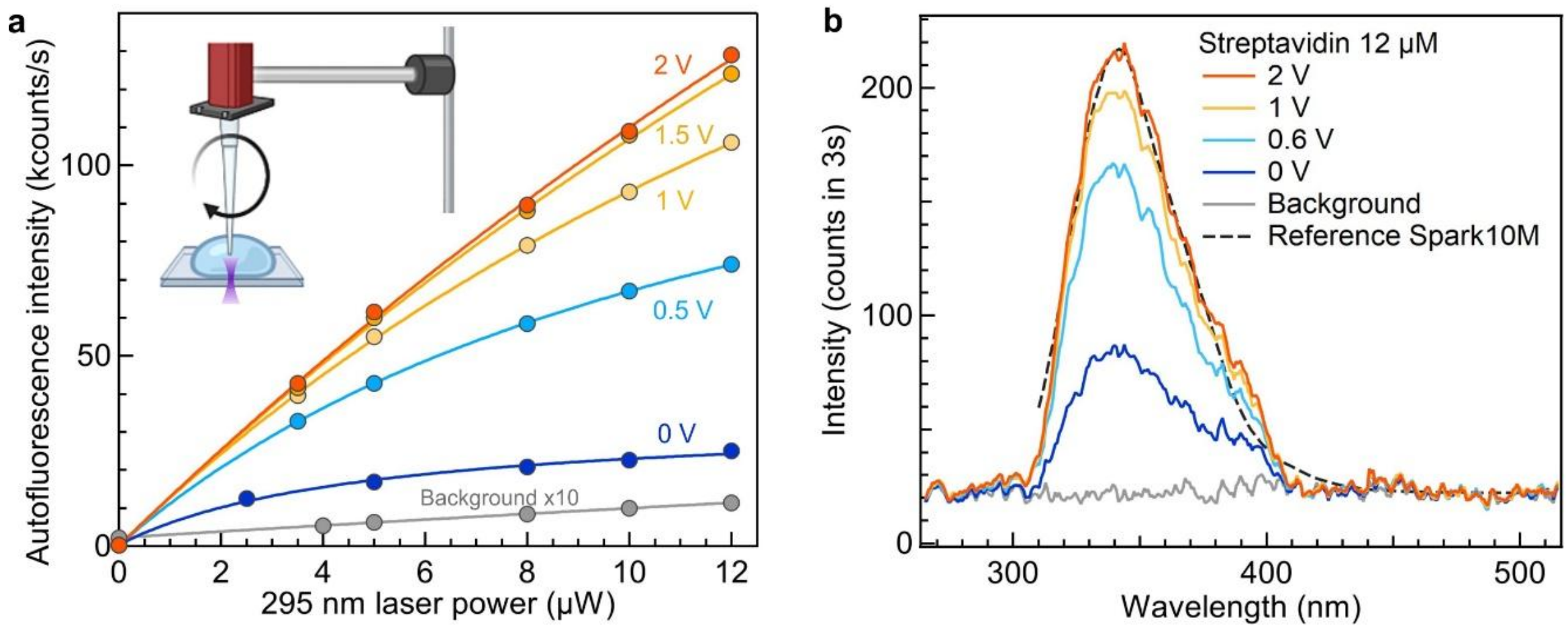

**Figure S1.** Throughout our experiments, we use a rotating stirring unit as depicted on the scheme in (a) to homogenize the solution and mitigate the UV photodamage of the proteins. The stirring speed is regulated by applying different voltages on the rotating motor. (a) Autofluorescence intensity recorded for a solution of 1 µM streptavidin as a function of the UV laser power with increasing stirring speeds from 0 V to 2 V. The background values (blank reference) are multiplied by 10 to ease viewing. At 6 µW laser power used for the rest of this work, the intensity remains nearly constant for stirring speeds above 1.4 V. We have decided to work at 1.8 V to give us a supplementary security margin and ensure optimum reproducibility between experiments. (b) Autofluorescence spectra recorded after the 80 µm confocal pinhole using an OceanOptics QEPro spectrometer with 100 µm entrance slit and 600 µm fiber core. The 295 nm laser power is 6 µW. The streptavidin concentration is 12 µM and we again use different voltages to control the stirring speed. The black dashed curve is the normalized reference spectrum acquired for the same streptavidin solution on a Tecan Spark 10M spectrofluorometer. At 0 V, we observe a lower autofluorescence intensity which we relate to the production of reactive oxygen species at the UV laser focus. Stirring the solution dilutes the UV-generated reactive oxygen species into the 300 µL solution and avoids observing the photodamage effects for voltages above 1.4V.

## S3. Matlab code used for numerical simulations

```
%function strepbiot(p)

p=1; % bias number to determine cooperativity. p=1 means fully random, p>1 is positive cooperativity, p<1 means negative cooperativity;

Nstrepini=1500; % Number of streptavidin molecules to run simulation, needs > 500 for proper convergence

NumTrial=50; % Number of binding trials for each molecule, needs > 15 for proper convergence

numpts=51; % Number of points desired in the biotin:streptavidin fraction between 0 and 5; 51 means 0.1 spacing between points

FracBiot=linspace(0,5,numpts); % Fraction of biotin:streptavidin, goes between 0 and 5 (enough to saturate the 4 binding sites)

Nbiot=FracBiot.*Nstrepini; % number of biotin molecules added

FracBoundResult=zeros(numpts,1); % initialisation result final bound fraction
%FracDef=zeros(numpts,1); % initialisation result defective fraction
%FracBoundtot=zeros(numpts,1); % initialisation result final total streptavidin for control

def=0; % defective rate in % of Streptavidin not having 4 active binding sites, 0 means perfect sample
wdef=[(100-def)/100 def/100]; % Probability weight normalized for defective Streptavidins

F0=100; % filtration efficiency in % of streptavidin with 0 biotin
F1=100; % filtration efficiency in % of streptavidin with 0 biotin
F2=100; % filtration efficiency in % of streptavidin with 0 biotin
F3=100; % filtration efficiency in % of streptavidin with 0 biotin

for i=1:numpts %loop for each biotin fraction

  Nstrep0=Nstrepini; % Streptavin with 0 biotin, initialisation at beginning all streptavidins have no biotin
  Nstrep1=0; % Streptavidin with 1 biotin
  Nstrep2=0; % Streptavidin with 2 biotins
  Nstrep3=0; % Streptavidin with 3 biotins
  Nstrep4=0; % Streptavidin with 4 biotins
  Ndef1=0; % Defective Streptavidins with only 1 active sites
  Ndef2=0; % Defective Streptavidins with only 2 active sites
  Ndef3=0; % Defective Streptavidins with only 3 active sites

  for j=1:Nbiot(i) % Loop for each biotin molecule within a specific fraction

  k=0;
  while k<NumTrial    % attempts for biotin binding

  if (Nstrep1+2*Nstrep2+3*Nstrep3+4*Nstrep4+Ndef1+2*Ndef2+3*Ndef3)<Nbiot(i)

  % Probabilities to bind to a streptavidin with 1, 2, .. binding sites are proportional to numbers to streptavidins, plus the cooperative parameter p

  % 3 different models to choose from, comment/uncomment the one you like

%       % Model 1 equal p among binding sites
%       P1=Nstrep0/Nstrepini;
%       P2=p*Nstrep1/Nstrepini;
```

```
%       P3=p*Nstrep2/Nstrepini;
%       P4=p*Nstrep3/Nstrepini;
%       P5=Nstrep4/Nstrepini;

%       % Model 2 additive sum p among binding sites
%       P1=Nstrep0/Nstrepini;
%       P2=p*Nstrep1/Nstrepini;
%       P3=(1+2*(p-1))*Nstrep2/Nstrepini;
%       P4=(1+3*(p-1))*Nstrep3/Nstrepini;
%       P5=Nstrep4/Nstrepini;

     % Model 3 multiplicative p among binding sites
     P1=Nstrep0/Nstrepini;
     P2=p*Nstrep1/Nstrepini;
     P3=p^2*Nstrep2/Nstrepini;
     P4=p^3*Nstrep3/Nstrepini;
     P5=(Nstrep4+Ndef1+Ndef2+Ndef3)/Nstrepini;

     wcalc=[P1 P2 P3 P4 P5]; % Defining probability weight
     w=wcalc./sum(wcalc);  % Probability weight normalized

     R = randsample('ABCDE',1,true,w); % Picking a random binding site or none

     if (Ndef1+Ndef2+Ndef3)<(Nstrepini*def/100)
       Rdef = randsample('OD',1,true,wdef); % Random choosing of defective Streptavidin or not
     else
       Rdef='O'; % If already found max defective streptavidins, stop testing
     end

     if (R=='A' && Rdef=='O')  % Binding to an active streptavidin with no bound biotin, binding is irreversible
       Nstrep1=Nstrep1+1;
       Nstrep0=Nstrep0-1;
       k=NumTrial;
     end
     if (R=='B' && Rdef=='O')  % Binding to a streptavidin with already 1 bound biotin
       Nstrep2=Nstrep2+1;
       Nstrep1=Nstrep1-1;
       k=NumTrial;
     end
     if (R=='C' && Rdef=='O')   % Binding to a streptavidin with already 2 bound biotins
       Nstrep3=Nstrep3+1;
       Nstrep2=Nstrep2-1;
       k=NumTrial;
     end
     if (R=='D')   % Binding to a streptavidin with already 3 bound biotins
       Nstrep4=Nstrep4+1;
       Nstrep3=Nstrep3-1;
       k=NumTrial;
     end

     if (R=='A' && Rdef=='D')  % Binding to an DEFECTIVE streptavidin with no bound biotin, binding is
irreversible
       Ndef1=Ndef1+1;
       Nstrep0=Nstrep0-1;
       k=NumTrial;
     end
     if (R=='B' && Rdef=='D')  % Binding to a DEFECTIVE streptavidin with already 1 bound biotin
```

```
      Ndef2=Ndef2+1;
      Nstrep1=Nstrep1-1;
      k=NumTrial;
    end
    if (R=='C' && Rdef=='D')   % Binding to a DEFECTIVE streptavidin with already 2 bound biotins
      Ndef3=Ndef3+1;
      Nstrep2=Nstrep2-1;
      k=NumTrial;
    end

    if R=='E'   % Finding a streptavidin with already 4 bound biotins, no effect, try again
      k=k+1;
    end
  end
  end

  end

  FracBoundResult(i)= ((1-F0/100).*Nstrep0+(1-F1/100).*Nstrep1+(1-F2/100).*Nstrep2+(1-
F3/100).*Nstrep3+Nstrep4+Ndef1+Ndef2+Ndef3)/Nstrepini; % Fully saturated Streptavidin fraction including
defective streptavidins
  %FracDef(i)= (Ndef1+Ndef2+Ndef3)/Nstrepini; % fraction of defective streptavidins (for control)
  %FracBoundtot(i)= (Nstrep1+2*Nstrep2+3*Nstrep3+4*Nstrep4)/(4*Nstrepini*(100-def)/100); % Total
fraction of bound biotins with no defect (for control)
end

plot(FracBiot,FracBoundResult)

%plot(FracBiot,FracBoundResult,FracBiot,FracBoundtot,FracBiot,FracDef) % for control, plots also total fraction
of bound biotins qnd defect rate

%end
```

<u>Additional comments:</u>

- The convergence of the numerical simulations is checked in Fig. S7 for the number of streptavidin molecules Nstrepini and the number of binding attempts NumTrial.
- An analytical expression is derived for the binding curve in Section S10 for the mutually-independent binding events (absence of cooperativity).

## S4. Equal, additive and multiplicative models used for cooperativity

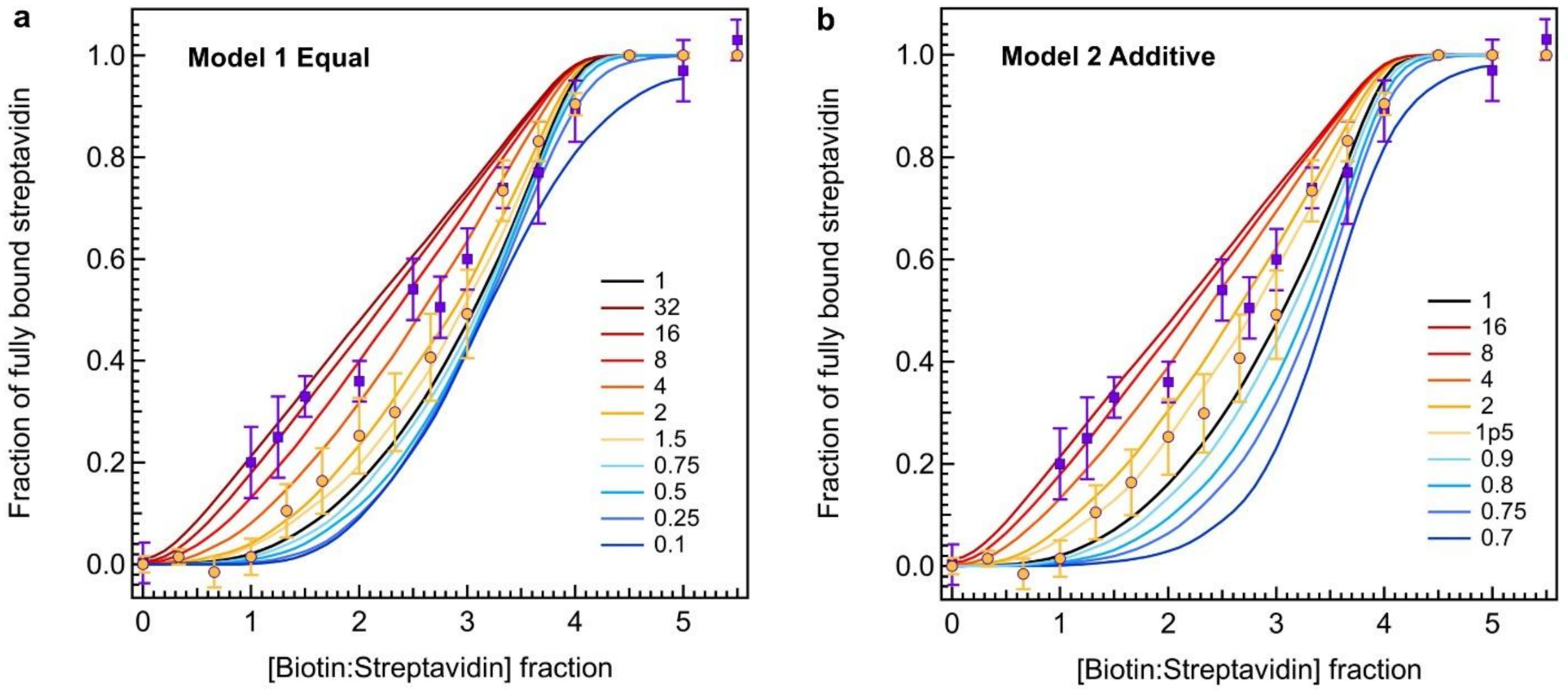

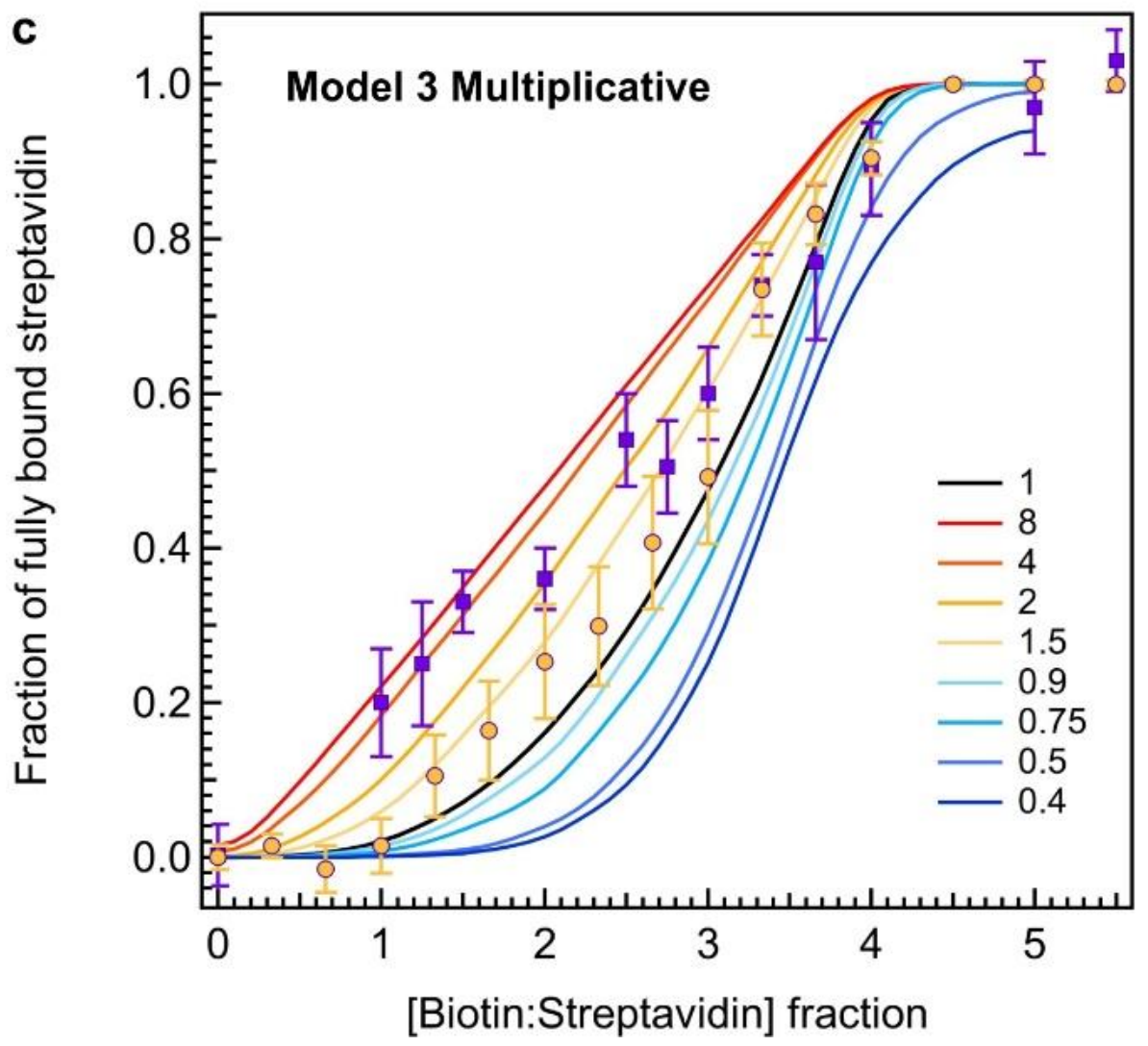

**d**

| | Probability weight for binding | | |
|---|---|---|---|
| **Number of biotins per streptavidin** | **Model 1 equal** | **Model 2 additive** | **Model 3 multiplicative** |
| 0 | $C_0$ | $C_0$ | $C_0$ |
| 1 | $p*C_1$ | $p*C_1$ | $p*C_1$ |
| 2 | $p*C_2$ | $(1+2(p-1))*C_2$ | $p^2*C_2$ |
| 3 | $p*C_3$ | $(1+3(p-1))*C_3$ | $p^3*C_3$ |
| 4 | $C_4$ | $C_4$ | $C_4$ |

**Figure S2.** Fraction of fully bound streptavidin as a function of the biotin to streptavidin relative concentration for different models and cooperativity parameters $p$ ($p > 1$ for positive cooperativity, $p < 1$ for negative cooperativity, and $p = 1$ for mutually independent binding events). The experimental markers reproduce the data in Fig. 2c. In (b,c), the affinities to bind to a specific streptavidin site vary with the number of filled binding sites. For the additive model in (b), the affinities cumulate: the affinity to bind to a streptavidin with 2 biotins is 2× higher than the affinity to bind to a streptavidin with 1 biotin. For the multiplicative model in (c), the affinities are multiplied: the affinity to bind to a streptavidin with 2 biotins is the square of the affinity to bind to a streptavidin with 1 biotin. The key equations are written in table (d). The Matlab code is given in Section S3. In all graphs, the black curve corresponds to the absence of cooperativity, orange colors indicate positive cooperativity and blue colors stand for negative cooperativity. (d) Definition of the probability weights used to describe cooperative binding with the different models. $C_i = N_i/N_{tot}$ describes the relative concentration of streptavidins bearing a number $i$ of biotins, $p$ is the cooperativity parameter. For mutually independent binding events and $p = 1$, all models yield identical results.

## S5. Control of filtration efficiency

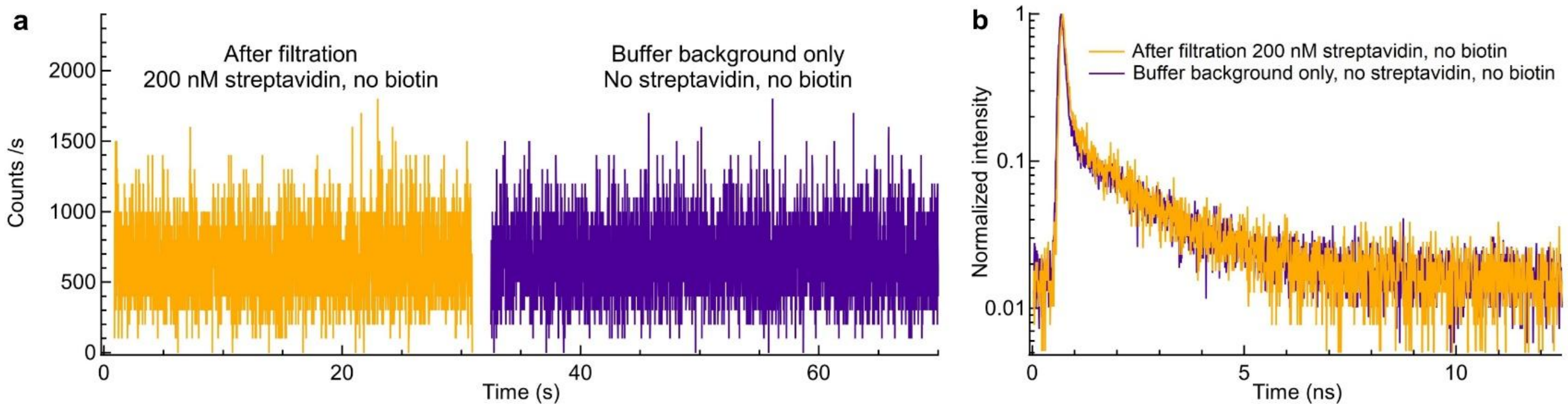

**Figure S3.** Demonstration that the biotin-coated silica bead solution is sufficient to filter all the streptavidin proteins at 200 nM concentration. For these experiments, no biotin solution was added to realize the most critical filtration test with a maximum of streptavidin proteins to be removed from the solution. After the filtration protocol with the biotin-coated silica beads, the detected autofluorescence intensity (a) is similar to the residual background level in presence of the buffer solution containing only 100 mM phosphate and 10 mM NaCl. The time-correlated single photon counting (TCSPC) data in (b) confirm the similarity between the autofluorescence intensity after filtration and the buffer background. This demonstrates that our filtration protocol is sufficient to remove all non-fully bound streptavidin proteins from the solution.

## S6. Influence of biotin on streptavidin UV autofluorescence

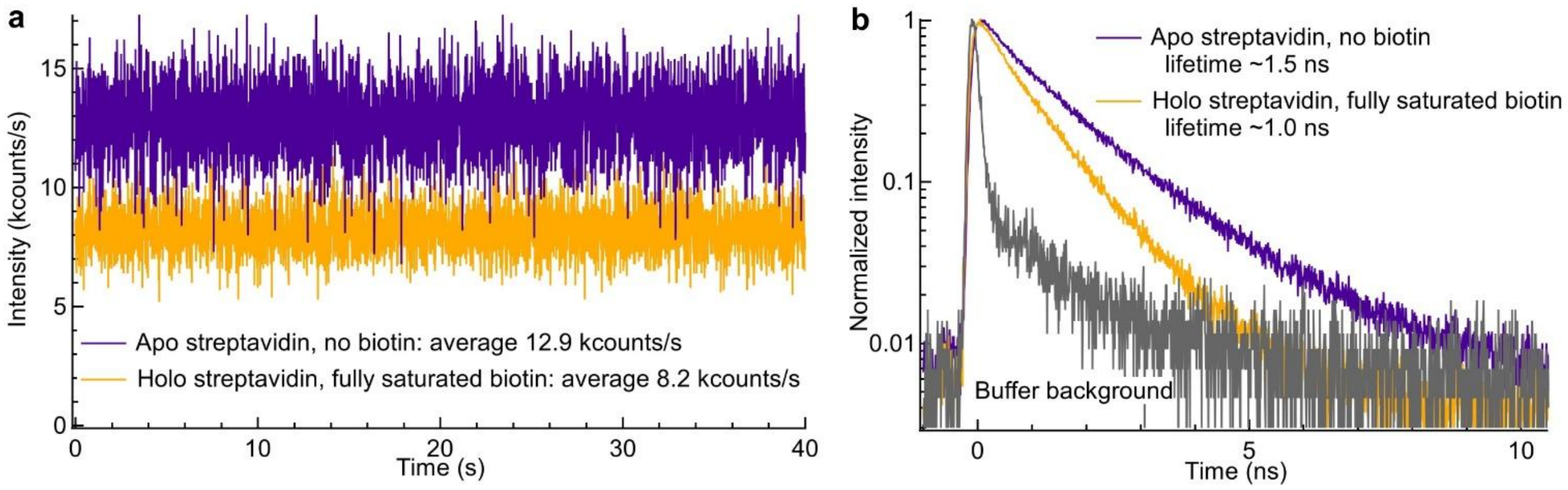

**Figure S4.** The presence of biotin fully saturating the streptavidin proteins has a moderate quenching influence on the autofluorescence emission, as seen by (a) a 36% decrease on the average intensity for a 200 nM streptavidin solution and (b) a reduction of the autofluorescence lifetime from 1.5 (apo-streptavidin) to 1.0 ns (holo-streptavidin). These findings are consistent with an earlier report using cuvette spectrofluorometer.[8] The biotin influence is fully taken into account in our analysis and does not change our conclusions. The time traces is (a) use a 10 ms binning time. The UV autofluorescence lifetime analysis uses a bi-exponential fitting with reconvolution from the instrument response function.[6] The autofluorescence lifetime expressed are intensity-averaged between the 2 individual exponential components.

## S7. Dimension of the interaction volume

To determine the effective radius used for the model in Fig. 2d, we have recorded the expansion of black ink droplets of known volumes pipetted into a water solution. The motivation behind these experiments is to visualize directly droplet dilution in aqueous solutions, in order to determine how the effective interaction volume scales with the droplet radius under realistic experimental conditions.

The pipetting is similar to the one used for biotin solutions using common laboratory equipment and process. The choice of black ink is motivated by the need for high optical contrast and good water solubility. We have checked that similar results can be obtained using Nile Blue solutions of 50 µM concentration. Typical images recorded 2 s after pipetting the droplet into the solution are shown in Fig. S5a. These images use a Thorlabs Zelux CMOS camera equipped with a 50 mm focal lens for quantitative imaging. Due to the black ink diffusion into the pure water solution, the radius of the effective interaction volume is larger than the radius R of the injected droplet volume, but does not evolve significantly over the 1-4 s time interval (Fig S5b). From the image set, we extract the effective radius of the interaction volume represented as a function of the radius R of the injected volume in Fig. S5c. The linear fit of the data determines the slope 1.5 used in the model for Fig. 2c-f.

The model assuming local concentration inhomogeneities in Fig. 2d considers that the interaction occurs within a spherical volume of radius 1.5 R + 1 mm, where R is the radius of the injected biotin volume (in mm). The slope of 1.5 could be determined experimentally from Fig. S5c. However, we empirically found that an additional +1 mm offset was needed in order to closely interpolate the experimental data for 25 µM biotin in Fig. 2c. While the exact physical origin of this 1 mm offset is unclear, the value seems plausible given the complex mixture dynamics and the simple approximation of the model.

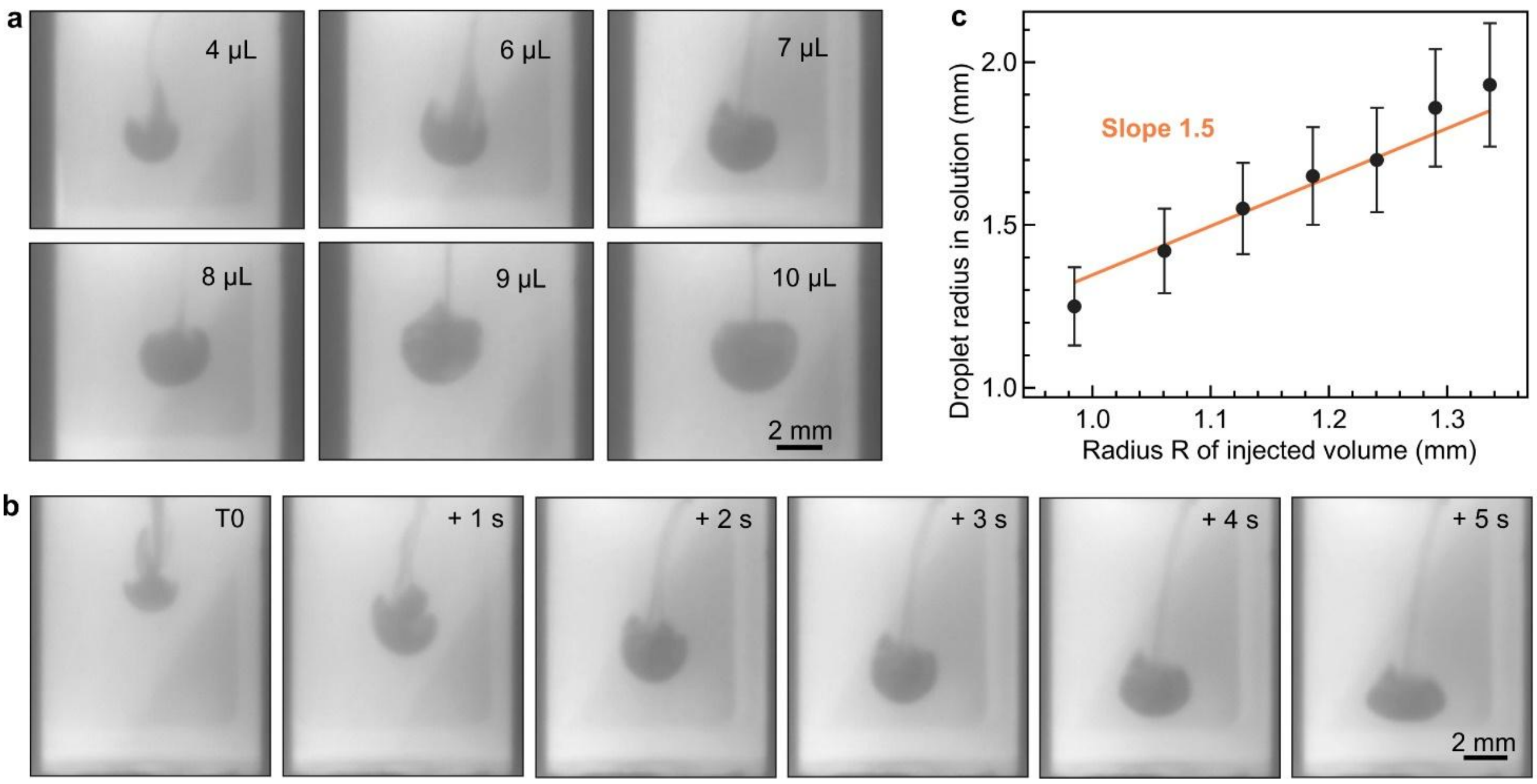

**Figure S5.** Determination of the effective radius used for the model in Fig. 2d. (a) Images of black ink droplets of calibrated volume injected into a water-filled cuvette. The images are recorded 2 s after injection with a Thorlabs Zelux CMOS camera equipped with a 50 mm focal lens. (b) Time lapse images for a 7 µL droplet in water. The dimension of the effective volume do not evolve significantly in the range 1-4 s. (c) Effective radius of the interaction volume determined from images similar to the ones in (a). The linear fit of the data determines the slope 1.5 used in the linear model for Fig. 2c-f. For each volume target, the experiments are repeated at least 5 times. The error bar correspond to twice the standard deviation. For non-spherical volumes, the equivalent radius was determined by averaging the horizontal and vertical dimensions, with respective weights 2/3 and 1/3 to account for the 3D shape.

## S8. Calibration of the UV confocal detection volume

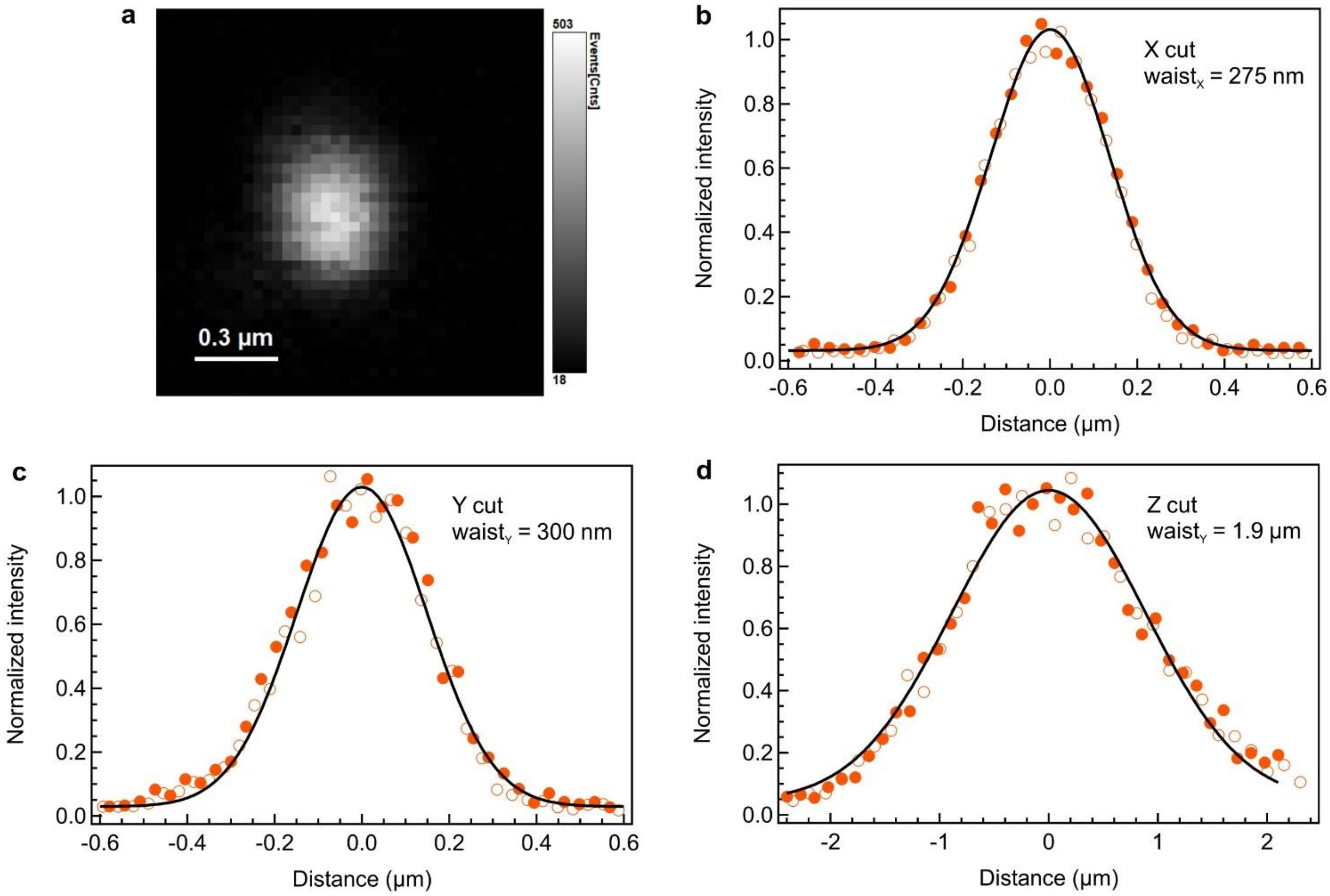

**Figure S6.** Calibration of the fluorescence microscope point spread function (PSF) by scanning a single nanohole of 90 nm diameter milled into a 100 nm thick opaque aluminum film filled with p-terphenyl UV-fluorescent molecules.[7] (a) Confocal transversal XY image of the nanohole. (b-d) Cross-cuts along the X, Y and Z directions together with numerical fits using a Gaussian function. Empty and filled markers represent two different experiment runs, yielding comparable results. The waists $w_{X,Y,Z}$ (radii at 1/e² intensity) are indicated on each graph. The confocal detection volume defined as $V_{det} = \pi^{3/2} w_X w_Y w_Z$ amounts to 0.9 ± 0.1 fL.[9]

## S9. Convergence check for the Monte Carlo simulations

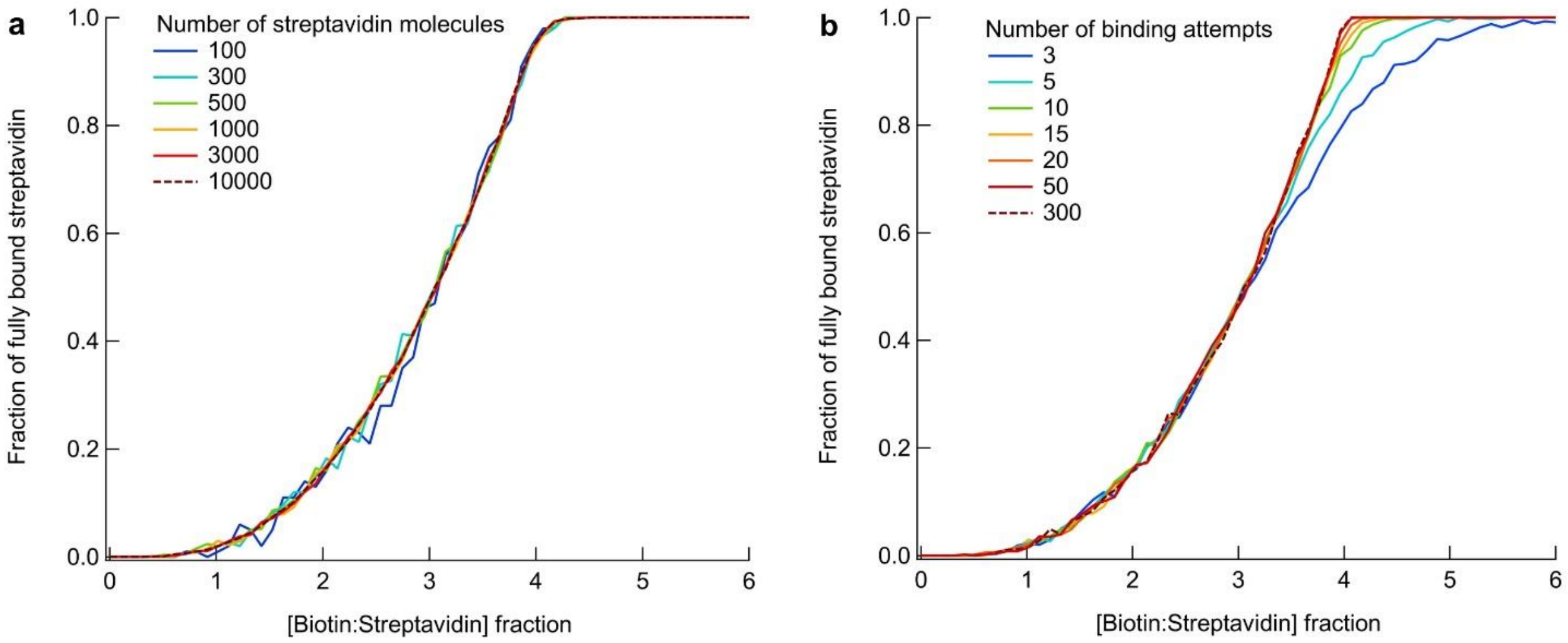

**Figure S7.** Checking the convergence of the numerical simulations used to compute the fraction of fully bound streptavidin as a function of the biotin to streptavidin relative concentration. Here we have considered the non-cooperative case and set $p = 1$ so that the model used (equal, additive or multiplicative) has no influence. In (a) we test the number of streptavidin molecules used in the model (Nstrepini in the Matlab code in Section S3). The calculations used in the rest of this work use Nstrepini = 1500 to ensure a good convergence. In (b) we look at the influence of the number of binding attempts (NumTrial in the Matlab code). The calculations used in the rest of this work use NumTrial = 50 to ensure a correct convergence at high biotin to streptatividin fractions.

## S10. Analytical expression for the mutually-independent binding model

Following the approach and notations introduced in the chapter 28 of ref.[10], the fraction of P molecules that are in the 4th-liganted state is $[PX_4]/(Q[P])$, where $[P]$ is the concentration of free molecules, $[X] = x$ is the concentration of free ligands and $Q$ is the so-called binding polynomial. For mutually independent binding events with affinity $K$, the binding polynomial can be expressed as $Q = (1 + Kx)^4$. Summing up on all possible pathways to bind 4 biotins, the concentration of fully-saturated streptavidin is $[PX_4] = 4!\ (Kx)^4[P]$, where the prefactor $4! = 24$ accounts for the different options to list 4 binding site choices in a specific order. Altogether, the fraction of fully bound streptavidin measured in our experiments can be expressed as $24\ (Kx/(1 + Kx))^4$. In this expression, $x$ denotes the biotin concentration. To adapt with the rest of the work which uses the biotin to streptavidin fraction, we introduce the variable replacement $x' = [X]/[S]$ and $K' = K\ [S]$ where $[S]$ is the total streptavidin concentration (constant in our work). With these new notations, the fraction of fully bound streptavidin molecules bearing 4 biotins becomes $24\ (K'x'/(1 + K'x'))^4$. This expression interpolates well with the Monte Carlo simulations in the absence of cooperativity shown in Fig. S7. Of course, this only applies to the interval $0 \leq x' \leq 4$.

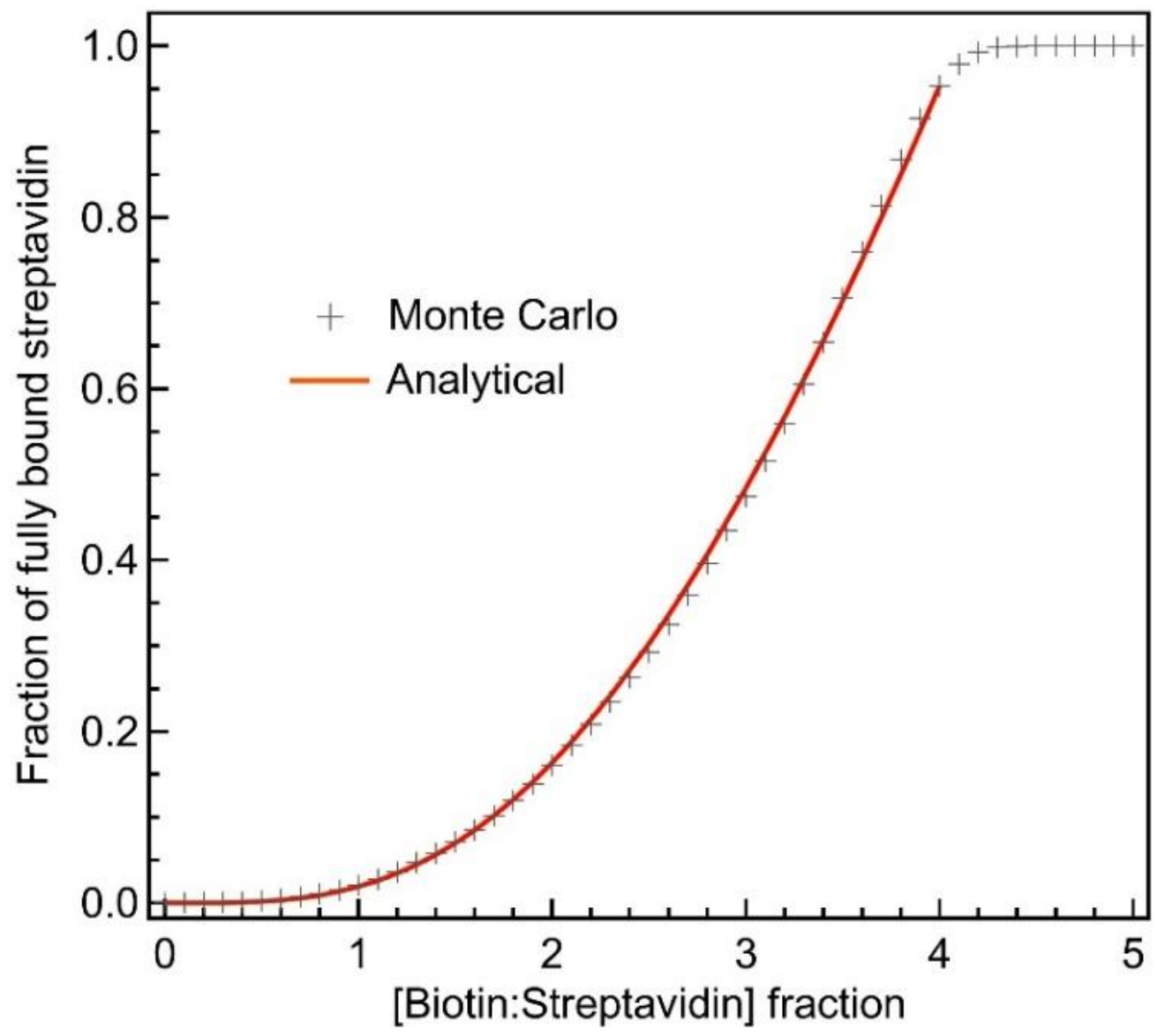

**Figure S8.** Comparison of the Monte Carlo results with the analytical expression $24\ (K'x'/(1 + K'x'))^4$ for the mutually-independent binding (absence of cooperativity). $x'$ is the biotin to streptavidin fraction (variable here) and $K' = K\ [S]$ is the binding affinity multiplied by the streptavidin concentration. Using $K'$ as a fitting parameter results in $K' = 0.2016\ \pm 0.0002$. The uncertainty is determined from the fitting process using Igor Pro v7. This value of $K'$comes very close to the value 0.2060 estimated by imposing the analytical expression to equal 1 for $x' = 4$.

## Supplementary References